\documentclass{revtex4}
\usepackage{graphicx}
\usepackage{fancyhdr}
\usepackage{amsmath}
\usepackage{color}
\usepackage{graphics}
\usepackage{amssymb}
\usepackage{latexsym}
\usepackage{supertabular}
\usepackage{slashed}
\usepackage{booktabs}
\usepackage{cases}
\usepackage{bigints}
\usepackage[abs]{overpic}

\newcommand{\nn}{\nonumber}
\newcommand{\simlt}{\lower.5ex\hbox{$\; \buildrel < \over \sim \;$}}
\newcommand{\simgt}{\lower.8ex\hbox{$\; \buildrel > \over \sim \;$}}
\newcommand{\resizeall}{\center\resizebox{0.6\textwidth}{!}}

\newcommand{\dsol}{\Delta_{21}}
\newcommand{\ddsol}{2\Delta_{21}}
\newcommand{\drct}{\Delta_{31}}

\newcommand{\ssol}{\sin^22\theta_{12}}
\newcommand{\srct}{\sin^22\theta_{13}}

\newcommand{\dms}{\Delta m^2_{21}}
\newcommand{\dmr}{\Delta m^2_{31}}
\newcommand{\dcT}{\Delta\chi^2}
\newcommand{\dcTm}{(\Delta\chi^2)_{\rm min}}
\newcommand{\cT}{\chi^2}

\newcommand{\exposure}{${\rm 20 \,GW_{th}}$$\cdot$5kt (12\% free-proton
weight fraction)$\cdot$5yrs}

\begin{document}

\title{Determination of mass hierarchy with medium baseline reactor
 neutrino experiments}

%

\author{Yoshitaro Takaesu}
\affiliation{Department of Physics and Astronomy, Seoul National
 University, Seoul 151-742, Korea}
\affiliation{School of Physics, KIAS, Seoul 130-722, Korea} 

\begin{abstract}
I discuss the sensitivity of future medium baseline
reactor antineutrino experiments on the neutrino mass hierarchy. By
 using the standard $\chi^2$ analysis, we find that the sensitivity depends
 strongly
on the baseline length $L$ and the energy resolution $(\delta E/E)^2 =
(a/\sqrt{E/{\rm MeV}})^2 + b^2$, where $a$ and $b$ parameterize the statistical and
systematic uncertainties, respectively. The optimal length is found to
 be $L \sim 40-55$ km, the larger resolution the shorter optimal $L$. 
For a 5 kton
 detector (with 12\% weight fraction of free proton) placed at $L \sim
 50$ km away
 from a $20 \,{\rm GW_{\rm th}}$ reactor, an experiment
 would determine the mass hierarchy with $\dcTm \sim 9$ on average after 5
 (15) or
 more years of running with the $(a,b)=(2,0.5)\%\,((3,0.5)\%)$ energy resolution.
This type of experiment can also
measure the relevant mixing parameters with the accuracy of $\sim 0.5\%$.
\end{abstract}

\maketitle

\thispagestyle{fancy}


\section{Introduction}
Now that a large $\theta_{13}$ has been measured at Daya Bay~\cite{An:2012eh,An:2012bu} and
RENO~\cite{Ahn:2012nd} experiments accurately, neutrino physics enters a
new era. One of the next challenges is determination of the mass
hierarchy. 

Among many ideas proposed, the medium baseline reactor antineutrino experiment~\cite{Petcov:2001sy,Choubey:2003qx,Learned:2006wy,Zhan:2008id,Batygov:2008ku,Zhan:2009rs,Ghoshal:2010wt} has 
stimulated various re-evaluations of its physics potential and sensitivity
recently. 
Some works utilize the Fourier transform
technique~\cite{Ciuffoli:2012iz,Ciuffoli:2012bs,Qian:2012xh}, first
discussed in
refs.~\cite{Learned:2006wy,Batygov:2008ku,Zhan:2008id}, to
distinguish the mass hierarchy. The main advantage of this technique is
that the mass hierarchy can be determined without precise knowledge of the
reactor antineutrino spectrum, the absolute value of the large mass-squared difference $|\Delta m_{31}^2|$, 
and the energy scale of a detector. Although interesting and attractive, 
this technique is somewhat subtle to incorporate the uncertainties of
the mixing parameters and to estimate its sensitivity to
the mass hierarchy. 

On the other hand, some works adopt the $\chi^2$
analysis~\cite{Ghoshal:2010wt,Ghoshal:2012ju,Qian:2012xh,Li:2013zyd} and new
measure based on Bayesian approach~\cite{Qian:2012zn}. These
methods utilize all available information from experiments, and it is
straightforward to incorporate the uncertainties to evaluate
the sensitivity, providing robust and complementary
results to the Fourier technique. 

In this proceedings, we analyze 
the sensitivity of medium baseline reactor antineutrino
experiments to the mass hierarchy for the baseline length of $10$--$100$ km and
the energy resolution $(\delta E/E)^2 =
\left(a/\sqrt{E/{\rm MeV}}\right)^2 +b^2$ in the range $2\% < a < 6\%$
and $b < 1\%$ with
the $\cT$ analysis. The optimal baseline length and the expected
statistical uncertainties of the neutrino parameters, $\ssol, \srct, \dms$
and $\dmr$, are also estimated. 

\section{Reactor antineutrino flux}
\label{basic}

In this section, we briefly discuss the evaluation of
how many electron antineutrinos, $\bar{\nu}_e$, would be detected at
a far detector with a medium 
baseline length from a reactor. 

In a nuclear reactor, antineutrinos are
mainly produced
via beta decay of the fission products of the four radio-active isotopes,
${}^{235}U, {}^{238}U, {}^{239}P_u$ and ${}^{241}P_u$, in the fuel.
%
The flux of antineutrinos with energy $E_{\nu}$ (MeV) at
a reactor of $P \,({\rm GW_{\rm th}})$ thermal power is then
expressed as~\cite{Zhan:2008id} 
\begin{align}
 \frac{dN}{dE_{\nu}} = \frac{P}{\sum_k f_k \epsilon_k}\phi(E_{\nu}) \times 6.24\times 10^{21},
\label{eq:flux}
\end{align}
where $f_k$ and $\epsilon_k$ are the
relative fission contribution and the released energy per fission of the
isotope $k$, respectively~\cite{Huber:2004xh}. $\phi(E_{\nu})$ is the
number of antineutrinos produced per fission~\cite{Vogel:1989iv}.

This rate is then modulated by neutrino oscillation. The $\bar{\nu}_e$ survival probability is expressed
as
\begin{align}
 P_{ee} &= \left|\sum_{i=1}^3 U_{ei} \exp\left(-i \frac{m_i^2 L}{2E_i} \right)
 U^*_{ei} \right|^2 \nn\\
= 1 &-\cos^4\theta_{13}\ssol\sin^2\left(\dsol\right) \nn\\
& -\srct\sin^2\left(|\drct|\right) \nn\\
  &
 -\sin^2\theta_{12}\srct\sin^2\left(\dsol\right)\cos\left(2|\drct|\right) \nn\\
& \pm \frac{\sin^2\theta_{12}}{2}\srct\sin\left(\ddsol\right)\sin\left(2|\drct|\right),
\label{pee2}
\end{align}
where $U_{ei}$ is the neutrino mixing-matrix element
relating the electron neutrino to the mass eigenstate ${\nu_i}$. The variables $m_i$ and
$E_i$ are the mass and energy of the corresponding mass eigenstate, 
while $\theta_{ij}$ represent the neutrino mixing angles. The oscillation
phases $\Delta_{ij}$ are defined as
\begin{equation}
 \Delta_{ij} \equiv \frac{\Delta m^2_{ij}L}{4E_{\nu}}, 
 \hspace{1em}(\Delta m^2_{ij} \equiv m^2_i -m^2_j)
\label{deltaij}
\end{equation}
with a baseline length $L$. 
The plus or minus sign in the last term of eq.~(\ref{pee2})
corresponds to the mass hierarchy; the plus sign is for normal hierarchy (NH), and
the minus sign for inverted
hierarchy (IH).
Note that this last term is the only source of the mass hierarchy difference.
We have neglected the matter effect because it is
small enough for the energy range and
 the baseline lengths we concern in this study~\cite{Hagiwara:2011kw}.
%
%

Similar as the current reactor experiments, such as Daya Bay~\cite{An:2012eh,An:2012bu},
RENO~\cite{Ahn:2012nd} and Double Chooz~\cite{Abe:2011fz},
future medium-baseline reactor-antineutrino experiments can also
use free protons as targets to detect electron antineutrinos via the inverse neutron-beta-decay (IBD) process, producing a neutron and a positron.
 The
 threshold neutrino energy of this process is 
\begin{equation}
E_{\rm thr}\sim
m_n -m_p +m_e. 
\end{equation}

The produced positron then interacts with scintillator, converting its
kinetic energy to photons. Eventually, the positron annihilates with an
electron in the detector and emits two 0.5 MeV photons. The
energies of photons are accumulated as the visible energy,
$E_{\rm vis}$, which is the sum of the positron's total energy and an
electron's rest energy,
\begin{align}
 E_{\rm vis} \sim& E_e +m_e 
 \sim ( E_{\nu} -0.8 ) \,{\rm MeV} .
\end{align} 

The observed
antineutrino distribution by a detector with $N_p$ free protons after an exposure time $T$ can
then be expressed as
\begin{align}
 \frac{dN}{dE^{\rm obs}_{\rm vis}} =&
 \frac{N_pT}{4\pi L^2}\int^\infty_{E_{\rm thr}} dE_{\nu}
 \frac{dN}{dE_{\nu}}
 P_{ee}(L,E_{\nu})\sigma_{\rm IBD}(E_{\nu}) \,G(E_{\nu}-0.8 {\rm MeV} -E^{\rm
 obs}_{\rm vis},\delta E_{\rm vis}),
\label{Nobs}
\end{align}
where $\sigma_{\rm IBD}(E_{\nu})$ is the cross section of the IBD
process~\cite{Vogel:1999zy}, $G$ is the detector response function
with the energy resolution
$\delta E_{\rm vis}$, and $E^{\rm obs}_{\rm vis}$ is the observed
visible energy by the detector. 
In this study, we take the normalized
gaussian function as the response function, i.e.,
\begin{align}
 G(E_{\rm vis}-E^{\rm obs}_{\rm vis},\delta E_{\rm vis}) = \frac{1}{\sqrt{2\pi}\delta
 E_{\rm vis}}\exp\left\{ -\frac{ \left(E_{\rm vis} -E^{\rm obs}_{\rm
 vis}\right)^2  }{ 2(\delta E_{\rm vis})^2
 } \right\}.
\quad
\label{eq:G}
\end{align} 
The detector energy resolution, $\delta E_{\rm vis}$, is parameterized into two parts,
 \begin{align}
  \frac{ \delta E_{\rm vis} }{E_{\rm vis}} = \sqrt{\left( \frac{ a }{
  \sqrt{E_{\rm vis}/{\rm MeV}} } \right)^2
  +b^2}.
 \label{eq:Eres}
 \end{align}
The first term in the square root represents the statistical uncertainty, and
the second one gives the systematic uncertainty~\cite{Eres}.  

\section{The sensitivity to the mass hierarchy}
\label{chi2}

After obtaining the energy distribution of reactor antineutrinos, we
would like to estimate the sensitivity of determining the mass hierarchy
using the standard $\cT$
analysis~\cite{Choubey:2003qx,Batygov:2008ku,Ghoshal:2010wt,Ghoshal:2012ju,Qian:2012xh,Li:2013zyd}. 

To set the stage, we introduce the $\cT$ function as
\begin{align}
 \cT = \cT_{\rm para} +\cT_{\rm stat}.
\label{eq:chi2_func}
\end{align}
The first term summarizes the prior knowledge on fitting parameters. In
reactor antineutrino experiments, these are the mixing angles, $\ssol$ and $\srct$, and the two mass-square
differences, $\dms$ and $|\dmr|$. In this study we also consider the
event-number normalization factor $f_{\rm sys}$, assuming the $3\%$
uncertainty. Their contributions look like,
\begin{align}
 \cT_{\rm para} &= \left\{ \frac{ (\ssol)^{\rm \,fit} -( \ssol )^{\rm \,input} }{ \delta\ssol }
 \right\}^2 +\left\{ \frac{ ( \srct )^{\rm \,fit} -(\srct)^{\rm \,input} }{ \delta\srct }
 \right\}^2 \nn\\
&+\left\{ \frac{ ( \dms )^{\rm \,fit} -( \dms )^{\rm \,input} }{
 \delta\dms }  \right\}^2 +\left\{ \frac{ ( |\dmr| )^{\rm \,fit} -(
 |\dmr| )^{\rm \,input} }{ \delta|\dmr| } \right\}^2  + \left( \frac{f_{\rm sys}^{\rm \,fit}-f_{\rm
 sys}^{\rm \,input} }{\delta f_{\rm sys}}  \right)^2.
\label{chi2para}
\end{align}
The input values $Y^{\rm input}$ and their uncertainties $\delta Y$ are listed in
Table~\ref{tb:fitting_params}.

\begin{table}
\caption{The input values $Y^{\rm input}$ and their uncertainties $\delta Y$ taken from refs.~\cite{Beringer:1900zz,An:2012eh,An:2012bu}. The uncertainty of $\srct$ can be
 5\% or less after 3 years running of Daya Bay experiment~\cite{Error_dmm31}.}
\begin{center}
\begin{tabular}{cccccc}
\hline\hline\addlinespace[2pt]
$Y$ & $\ssol$ & $\srct$ & $\dms \,{\rm eV}^2$ & $|\dmr| \,{\rm eV}^2$ &
 $f_{\rm sys}$ \\[2pt]
\hline \addlinespace[2pt]
$Y^{\rm input}$ & $0.857$ & $0.089$ & $7.50\times 10^{-5}$ & $2.32\times 10^{-3}$
		 & $1$\\
$\delta Y$ & $0.024$ & $0.005$ & $0.20\times 10^{-5}$ & $0.1\times
		 10^{-3}$ & 0.03 \\
\hline\hline
\end{tabular}
\end{center}
\label{tb:fitting_params}
\end{table}

The second term in (\ref{eq:chi2_func}) represents 
the statistical fluctuation. When we introduce binning w.r.t. $E^{\rm
obs}_{\rm vis}$, 
it looks like
\begin{align}
\cT_{\rm stat} = \sum_i \left( \frac{ N_i^{\rm \,fit} -N_i^{\rm NH(IH)} }{
\sqrt{N_i^{\rm NH(IH)}} }  \right)^2
\label{chi2stat}
\end{align}
with the summation running over all the bins. 
Here, $N_i^{\rm NH(IH)}$ is the event number for the $i_{\rm th}$ bin when 
the hierarchy is NH
(IH), while $ N_i^{\rm fit}$ is the theoretical prediction of the event
number either with right or wrong mass hierarchy, calculated as
a function of the four model parameters and the normalization factor
$f_{\rm sys}$, which are all varied under the
constraints of (\ref{chi2para}). In this study we
prepare the data $N_i^{\rm NH(IH)}$ by using eq.~(\ref{Nobs}) with the
input values of the five parameters for each mass hierarchy. 

In the limit of
infinitely many events, the bin size can be reduced to zero, and the sum
(\ref{chi2stat}) can be replaced by an integral.
%
%
Although a finite bin size is required for actual experiments, we
adopt this zero-bin-size limit as a measure of the maximum sensitivity. 

We then define $\dcT$ as
\begin{align}
 \dcT = \cT -\cT_{\rm min},
\label{eq:dcT}
\end{align}
where $\chi^2_{\rm min}$ is the minimum of $\chi^2$, which is obviously zero in our approximation of neglecting
statistical fluctuations in data, $N_i^{\rm NH(IH)}$. When wrong mass hierarchy is
assumed in the fit, the minimum of $\Delta\chi^2$, $(\Delta\chi^2)_{\rm min}$, will deviate from zero, and
the wrong mass hierarchy can be rejected with the significance $\sqrt{(\Delta\chi^2)_{\rm min}}$.

\clearpage
\section{Results}
\label{sec:results}
In this section, we discuss the sensitivity to the mass hierarchy, the optimal
length and the statistical uncertainties of the neutrino parameters.
All our results are obtained by assuming a reactor of 20 ${\rm GW_{th}}$
thermal power, a far detector of 5 kton fiducial
volume with 12\%
weight fraction of free proton and 5 years
exposure time. 


\begin{figure}[thpb]
\center\resizebox{1.0\textwidth}{!}{
\includegraphics{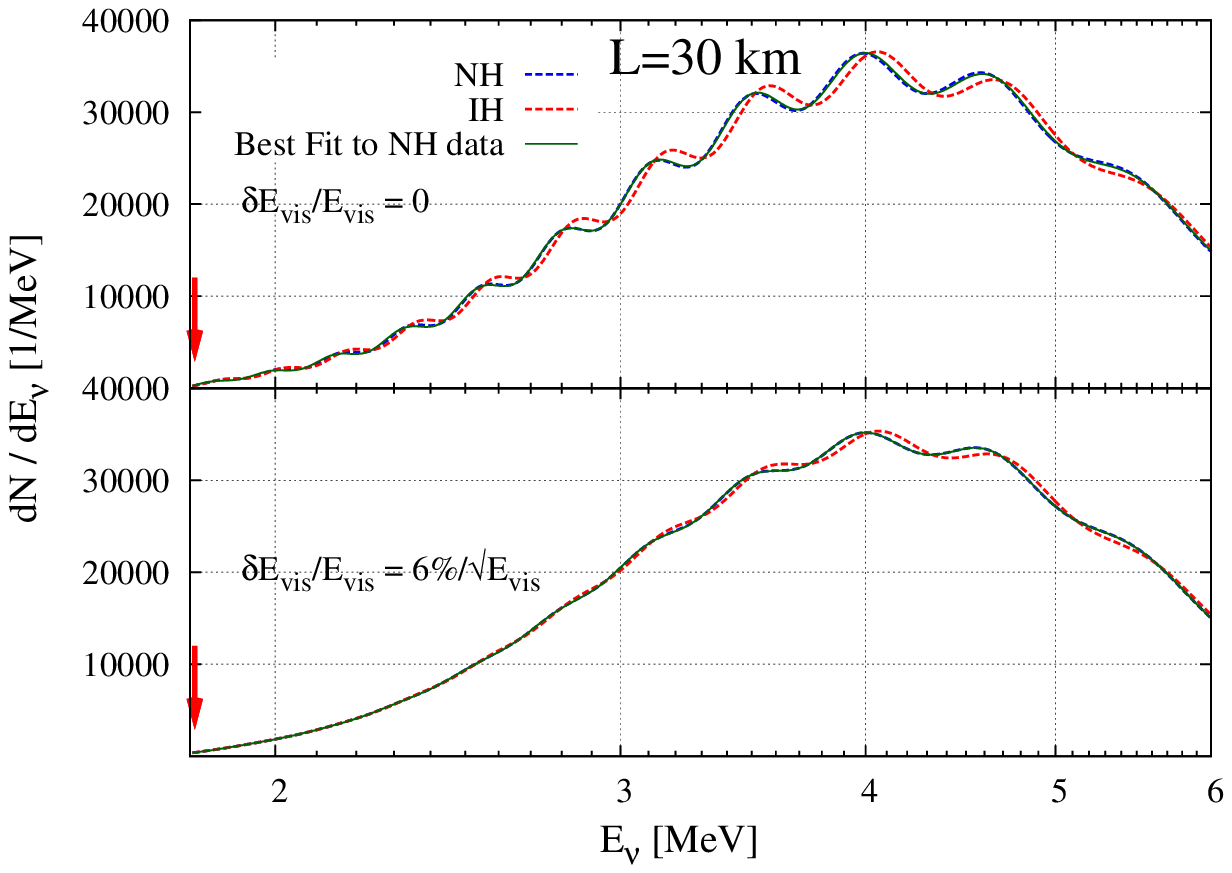}
\includegraphics{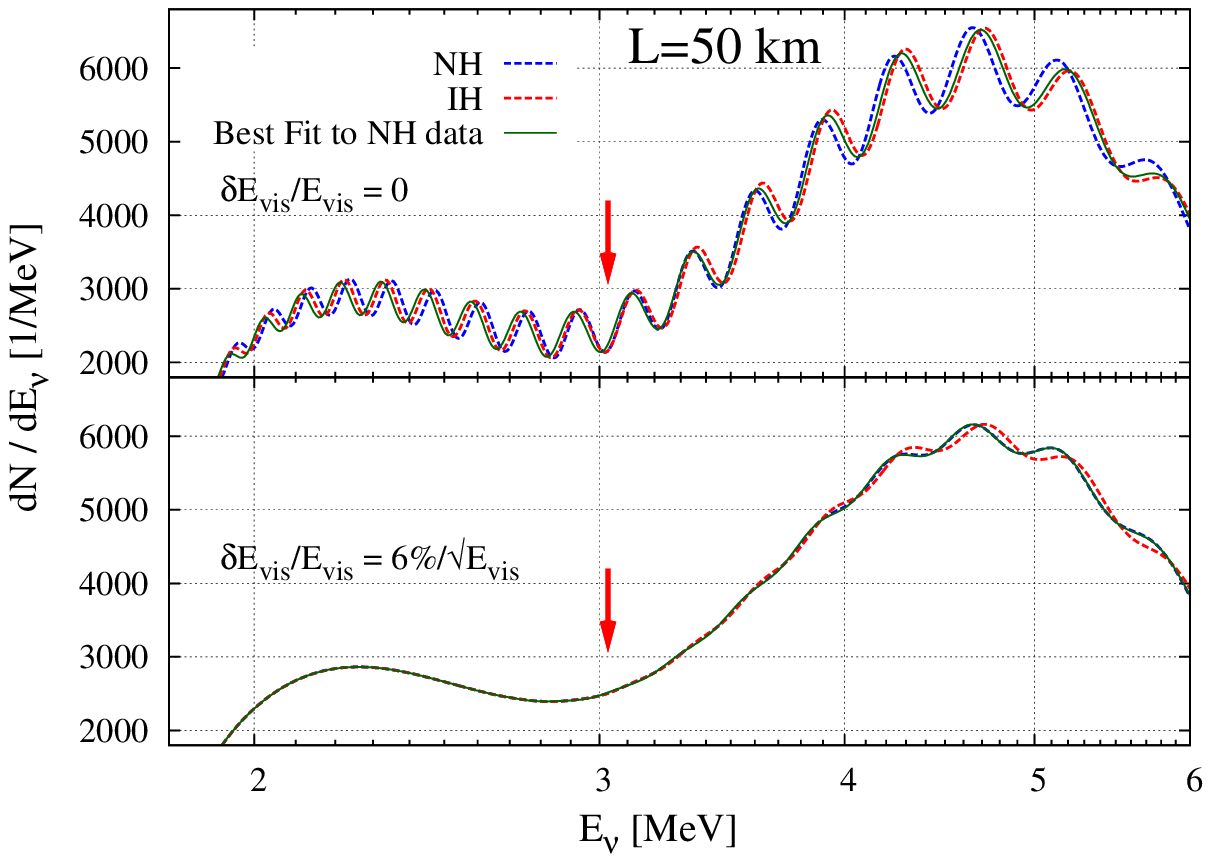}
}
\caption{The energy distribution of reactor antineutrinos
  with \exposure\, exposure and the baseline length $L = 30$ km (left) and
 $50$ km (right).
{\bf Upper}: The case with exact $E_{\nu}$ measurement, where the dashed blue and dashed red curves are
 for NH and IH,
 respectively. The solid curve shows the
 best fit of IH assumption to the NH data.
 The red arrow points out
 the energy at which the difference due to
 mass hierarchy vanishes.
{\bf Lower}: $6/\sqrt{E_{\rm vis}}\,\%$ energy resolution case.}
\label{fig:EventDistmin_fit2nh_combine_30}
\end{figure}

Figures~\ref{fig:EventDistmin_fit2nh_combine_30} shows energy
distributions for $L = 30$ km (left) and 50 km (right), 
in which the exact $E_{\nu}$ measurement is assumed in the upper panel, whereas in the lower
panel the energy
resolution of $a = 6\%$ with $b = 0$ in eq.~(\ref{eq:Eres}) is assumed. The dashed blue curve corresponds to
the NH case, and the dashed red curve to the IH case,
while the solid curve is obtained using the parameter values fitted to
the NH data with the ``wrong'' IH assumption. The red arrow points out
the energy at which the difference due to mass hierarchy vanishes (the
degeneracy point). 

At $L = 30$ km, the solid curve
almost coincides with the dashed blue one even with the exact energy
measurement, implying that it is almost impossible to distinguish
the mass hierarchy by experiments at $L =  30$ km. This is because the
small phase shift between the NH and IH predictions can be absorbed by a
small shift in $|\dmr|$ by a fraction of its present uncertainty, $0.1
\times 10^{-3} {\rm eV}^2$.

The situation changes when the second peak of the mass-hierarchy dependent term appears in
the energy range. The mass hierarchy difference can no longer be
absorbed by a shift in $|\dmr|$ since the relative phase difference
between the NH and IH oscillations changes across the degeneracy point.
There is no way to make the differences on the both sides compensated, 
resulting in the distinct mismatch between the dashed blue curve (the NH
data) and the solid curve (the best fit under the IH
assumption) as shown in the upper panel of the right plot in
Fig.~\ref{fig:EventDistmin_fit2nh_combine_30}, where the antineutrino energy
is exactly measured. 

Once the finite energy resolution is
introduced, the phase difference in the lower energy side of the
degeneracy point is significantly smeared out as it oscillates faster w.r.t. 
$E_{\nu}$ at the low energy. Hence it is easier for one oscillation
period to be covered by a sizable Gaussian profile of the detector response function. The remaining difference in the
higher energy side can then be absorbed by a small shift in $|\dmr|$,
resulting in an excellent fit (solid curve) to the NH data (blue
dashed curve) in the lower panel of the right plot in
Fig.~\ref{fig:EventDistmin_fit2nh_combine_30}, shown for
$6\%/\sqrt{E/{\rm MeV}}$
energy resolution. 

\begin{figure}[thpb]
\center\resizebox{1.0\textwidth}{!}{
 \includegraphics{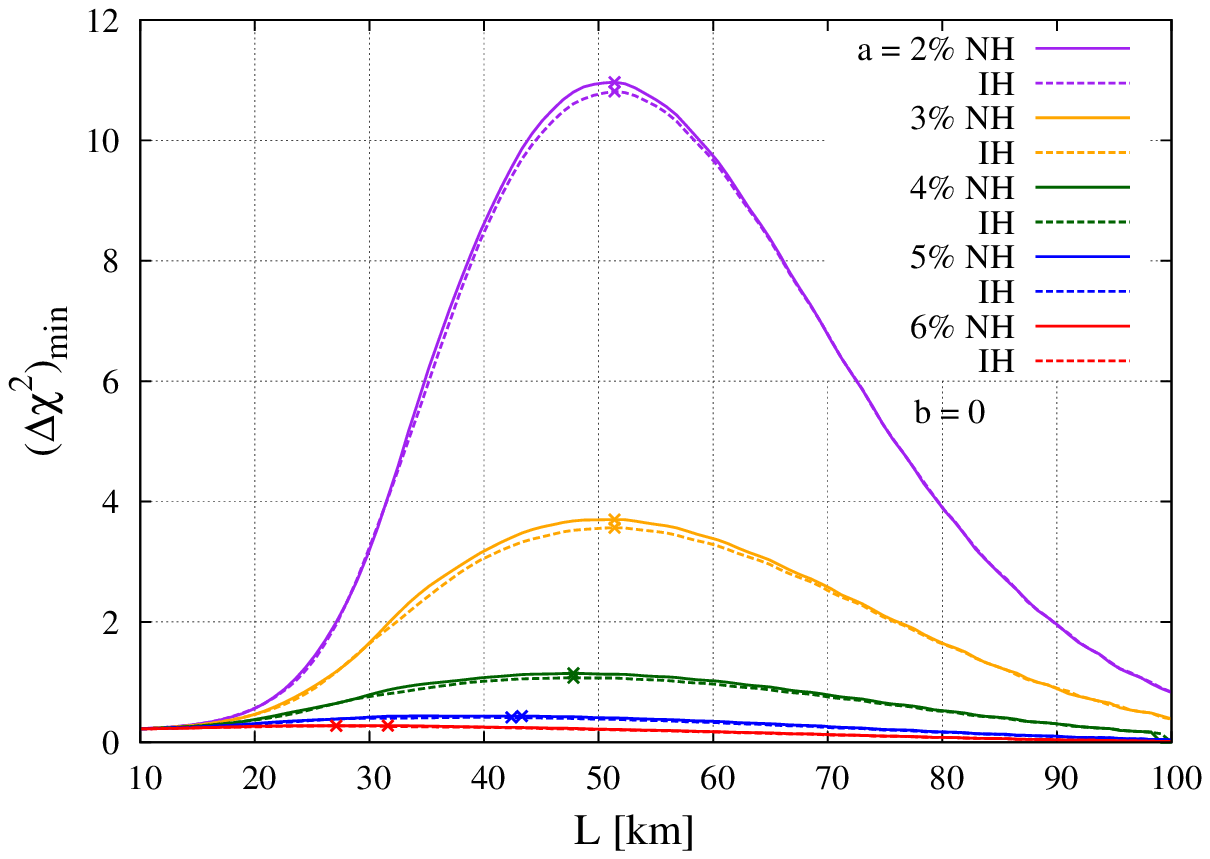}
 \includegraphics{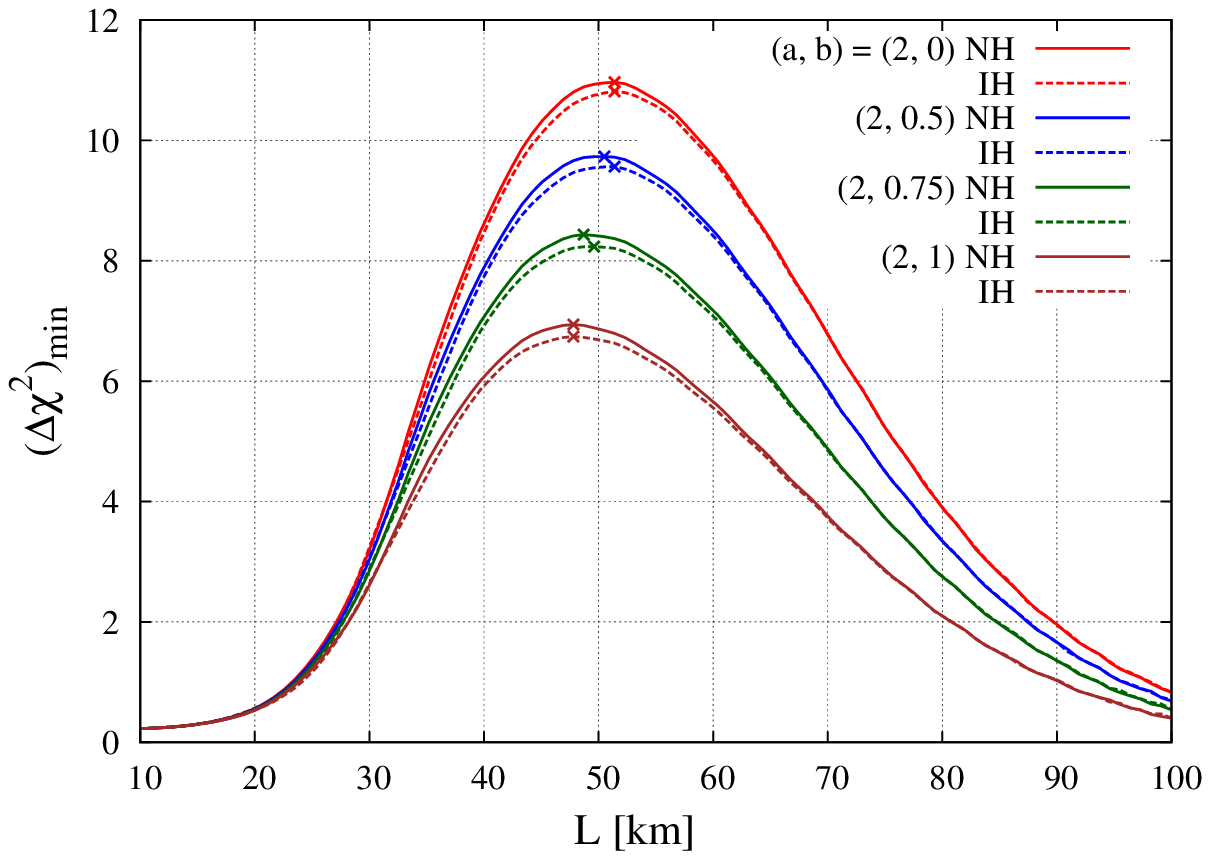}
 }
 \caption{$(\dcT)_{\rm min}$ for mass hierarchy discrimination shown as a function of the baseline
  length $L$ for \exposure\, exposure. The solid curves are for the NH cases
  and dashed curves for the IH cases. The cross symbols
  mark the optimal baseline lengths. {\bf Left}: The energy resolution in
  eq.~(\ref{eq:Eres}) is varied with $a=2$ to 6\% and $b=0$, from the top to the
  bottom.  {\bf Right}: The energy resolution is varied with 
   $a = 2\%$ and $b = 0\%, 0,5\%, 0.75\%, 1\%$, 
   from the top to the bottom.} 
 \label{fig:dchi2_combine}
 \end{figure}

The left plot in Fig.~\ref{fig:dchi2_combine} shows the resulted
$(\Delta\chi^2)_{\rm min}$ value as a function of the baseline length $L$,
for several energy resolutions, $a = 2,3,4,5$ and $6\%$ (with $b=0$) in
eq.~(\ref{eq:Eres}), from the top to the bottom.
Solid curves are for NH, while dashed curves are for IH. The results
clearly show that the mass hierarchy can be determined by those
experiments only if the energy resolution of the detector is
$3\%/\sqrt{E/{\rm MeV}}$ or better, and that the optimal baseline length
(as shown by the cross symbol) is around 50 km for that
resolution. The small $(\dcT)_{\rm min}$ for the baseline
length $L < 40$ km and $L>80$ km is due to a shift in $|\dmr|$ and low
statistics, respectively. For the $a=5$ and $6\%$ cases
$(\Delta\chi^2)_{\rm min}$ stays almost zero at all $L$. 

The right plot in Fig.~\ref{fig:dchi2_combine} shows the
$(\dcT)_{\rm min}$
value as a function of the baseline length $L$ for different $b$ values
with $a = 2\%$.
 The curves from the top to the bottom are obtained for $b = 0\%, 0.5\%,
 0.75\%$ and $1\%$, respectively. The effect of the systematic uncertainty is
significant as discussed in ref.~\cite{Qian:2012xh},
reducing the peak value of $(\dcT)_{\rm min}$ from 11.0~($b=0$) to
 9.7~($b=0.5\%$), 8.4~($b=0.75\%$) and 6.9~($b=1\%$) for NH. 
The optimal $L$ shortens
from 51 km for $(a,b) = (2,0)\%$ to 47 km for $(a,b) = (2,1)\%$. 

\begin{figure}[thpb]
\resizeall{
\includegraphics{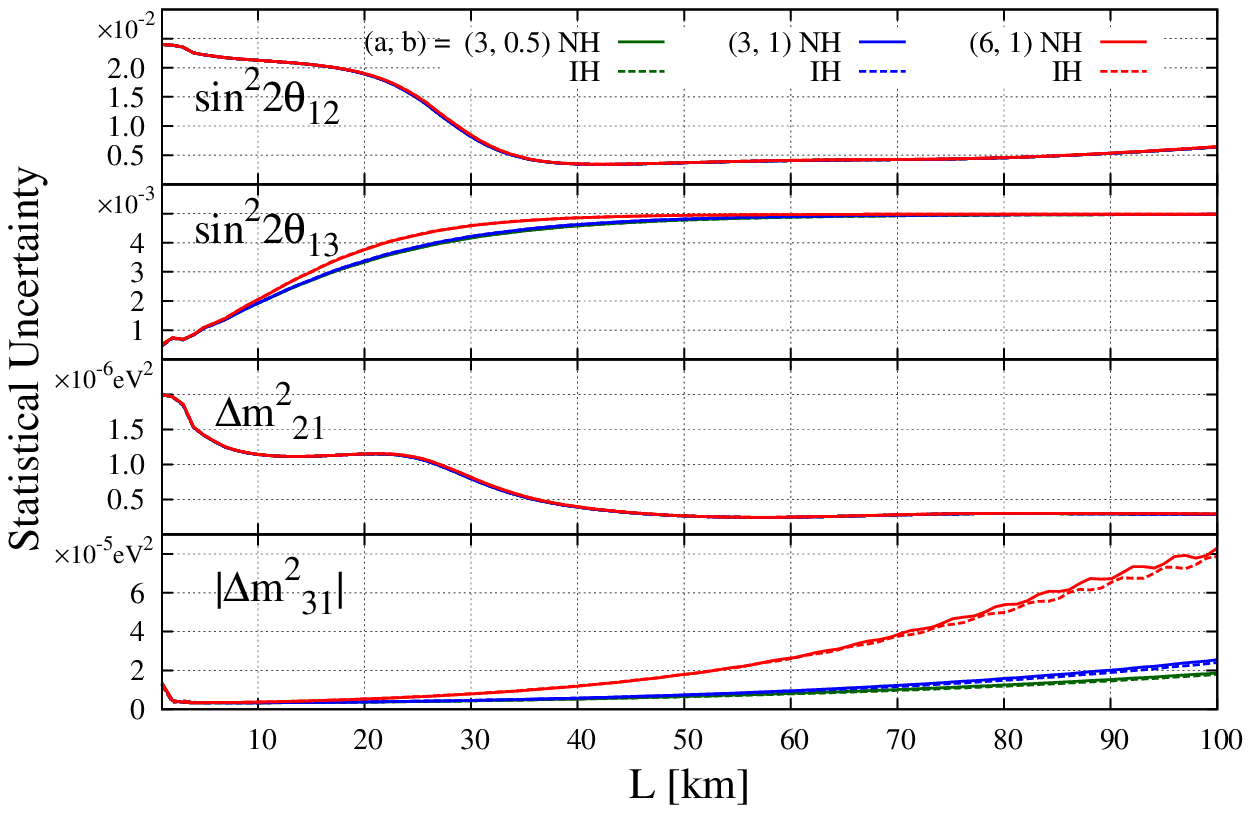}
}
\caption{The statistical uncertainties of the neutrino model parameters measured
 by this experiment as functions of the baseline length $L$ after
 \exposure\, exposure. The results are shown for both hierarchy (NH by solid and IH by dashed
 curves) and for the energy resolution of eq.~(\ref{eq:Eres}) with $(a,b)=(3,0.5),(3,1)$ and
 $(6,1)\%$.}
\label{fig:param_errors}
\end{figure}

In addition, the neutrino parameters, $\ssol, \dms$ and $|\dmr|$, can be measured accurately
with statistical uncertainties shown in
Fig.~\ref{fig:param_errors}. 
We find
\begin{subequations}
\begin{align}
\delta\ssol \sim& 4\times 10^{-3} \,(0.5\%), \\
\delta \dms \sim& 3\times 10^{-7} {\rm eV^2} \,(0.4\%), \\
\delta |\dmr| \sim& 7\times 10^{-6} {\rm eV^2} \,(0.3\%),
\end{align}
\label{eq:param_errors}
\end{subequations}
%
%
with the energy resolution of $(a, b) = (3,0.5)\%$ at $L = 50$
km; the percentage values in the parentheses denote the relative
accuracy of the measurement. Those uncertainties are almost independent of the mass hierarchy and of
the energy resolution, with
 the only exception of the $|\dmr|$ uncertainty for which the larger resolution
 results in the larger uncertainty.
$\srct$
 and $|\dmr|$ are measured most accurately around $L\sim 1$ km.
%
%

\section{Discussions and Conclusion}
\label{conclusion}
In this proceedings we have investigated the sensitivity of 
medium-baseline reactor-electron-antineutrino oscillation experiments for
determining the neutrino mass hierarchy by performing the standard
$\cT$ analysis.

We have carefully studied the impacts of the energy resolution 
$(\delta E/E)^2 = \left(a/\sqrt{E/{\rm MeV}}\right)^2 +b^2$
and find that the sensitivity strongly depends on it.
 The optimal baseline length
 is
found to depend slightly on the energy resolution, preferring the length
slightly shorter than 50 km for the energy resolution of
$(a,b)=(2,0.75)\%$ and $(2,1)\%$.
At the optimal baseline length,
the energy resolution better than the
$3\%/\sqrt{E/{\rm MeV}}$ level is needed to determine the neutrino mass
hierarchy pattern. 
For a 5 kton
 detector (with 12\% weight fraction of free proton) placed at $L \sim
 50$ km away
 from a $20 \,{\rm GW_{\rm th}}$ reactor, an experiment would determine the
 mass hierarchy
 with $\dcT\sim 9$ on average after five or more years of running if the
 energy resolution of
$(a,b)=(2,0.5)\%$ is achieved, while a
factor of three larger or longer experiment is needed to achieve the
same goal for the energy resolution of $(a,b)=(3,0.5)\%$.


It is also found that this experiment can measure the neutrino
parameters, $\ssol$, $\dms$ and $|\dmr|$, very accurately as shown in (\ref{eq:param_errors}) for an
experiment of \exposure\, at
$L\sim 50$ km.


\begin{acknowledgments}
This proceedings is based on the work in collaboration with Shao-Feng
 Ge, Kaoru Hagiwara and Naotoshi Okamura. We wish to thank Jun Cao,
 Jarah Evslin, Soo-Bong Kim, Serguey Petcov, Xin Qian, Yifang Wang and Xinmin Zhang for
valuable discussions on reactor neutrino experiments. 
This work
was in part supported by Korea Neutrino Research Center (KNRC) through National
Research Foundation of Korea Grant.
\end{acknowledgments}

\bigskip 

\bibliographystyle{bib2/general}
\bibliography{bib2/reference,bib2/neutrino}

\begin{thebibliography}{24}%
\makeatletter
\providecommand \@ifxundefined [1]{%
 \@ifx{#1\undefined}
}%
\providecommand \@ifnum [1]{%
 \ifnum #1\expandafter \@firstoftwo
 \else \expandafter \@secondoftwo
 \fi
}%
\providecommand \@ifx [1]{%
 \ifx #1\expandafter \@firstoftwo
 \else \expandafter \@secondoftwo
 \fi
}%
\providecommand \natexlab [1]{#1}%
\providecommand \enquote  [1]{``#1''}%
\providecommand \bibnamefont  [1]{#1}%
\providecommand \bibfnamefont [1]{#1}%
\providecommand \citenamefont [1]{#1}%
\providecommand \href@noop [0]{\@secondoftwo}%
\providecommand \href [0]{\begingroup \@sanitize@url \@href}%
\providecommand \@href[1]{\@@startlink{#1}\@@href}%
\providecommand \@@href[1]{\endgroup#1\@@endlink}%
\providecommand \@sanitize@url [0]{\catcode `\\12\catcode `\$12\catcode
  `\&12\catcode `\#12\catcode `\^12\catcode `\_12\catcode `\%12\relax}%
\providecommand \@@startlink[1]{}%
\providecommand \@@endlink[0]{}%
\providecommand \url  [0]{\begingroup\@sanitize@url \@url }%
\providecommand \@url [1]{\endgroup\@href {#1}{\urlprefix }}%
\providecommand \urlprefix  [0]{URL }%
\providecommand \Eprint [0]{\href }%
\providecommand \doibase [0]{http://dx.doi.org/}%
\providecommand \selectlanguage [0]{\@gobble}%
\providecommand \bibinfo  [0]{\@secondoftwo}%
\providecommand \bibfield  [0]{\@secondoftwo}%
\providecommand \translation [1]{[#1]}%
\providecommand \BibitemOpen [0]{}%
\providecommand \bibitemStop [0]{}%
\providecommand \bibitemNoStop [0]{.\EOS\space}%
\providecommand \EOS [0]{\spacefactor3000\relax}%
\providecommand \BibitemShut  [1]{\csname bibitem#1\endcsname}%
\let\auto@bib@innerbib\@empty
\bibitem [{\citenamefont {An}\ \emph {et~al.}(2012)\citenamefont {An} \emph
  {et~al.}}]{An:2012eh}%
  \BibitemOpen
  \bibfield  {author} {\bibinfo {author} {\bibfnamefont {F.~P.}\ \bibnamefont
  {An}} \emph {et~al.} (\bibinfo {collaboration} {Daya Bay Collaboration}),\
  }\bibfield  {booktitle} {\href {\doibase 10.1103/PhysRevLett.108.171803}
  {}\bibfield  {journal} {\bibinfo  {journal} {Phys.Rev.Lett.}\ }\textbf
  {\bibinfo {volume} {108}\ }(\bibinfo {year} {2012})\ \bibinfo {pages}
  {171803} [\Eprint {http://arxiv.org/abs/1203.1669}
  {arXiv:1203.1669}]}\BibitemShut {NoStop}%
\bibitem [{\citenamefont {An}\ \emph {et~al.}(2013)\citenamefont {An} \emph
  {et~al.}}]{An:2012bu}%
  \BibitemOpen
  \bibfield  {author} {\bibinfo {author} {\bibfnamefont {F.}~\bibnamefont {An}}
  \emph {et~al.} (\bibinfo {collaboration} {Daya Bay Collaboration}),\
  }\bibfield  {booktitle} {\href {\doibase 10.1088/1674-1137/37/1/011001}
  {}\bibfield  {journal} {\bibinfo  {journal} {Chin. Phys.}\ }\textbf {\bibinfo
  {volume} {C37}\ }(\bibinfo {year} {2013})\ \bibinfo {pages} {011001} [\Eprint
  {http://arxiv.org/abs/1210.6327} {arXiv:1210.6327}]}\BibitemShut {NoStop}%
\bibitem [{\citenamefont {Ahn}\ \emph {et~al.}(2012)\citenamefont {Ahn} \emph
  {et~al.}}]{Ahn:2012nd}%
  \BibitemOpen
  \bibfield  {author} {\bibinfo {author} {\bibfnamefont {J.~K.}\ \bibnamefont
  {Ahn}} \emph {et~al.} (\bibinfo {collaboration} {RENO collaboration}),\
  }\bibfield  {booktitle} {\href {\doibase 10.1103/PhysRevLett.108.191802}
  {}\bibfield  {journal} {\bibinfo  {journal} {Phys.Rev.Lett.}\ }\textbf
  {\bibinfo {volume} {108}\ }(\bibinfo {year} {2012})\ \bibinfo {pages}
  {191802} [\Eprint {http://arxiv.org/abs/1204.0626}
  {arXiv:1204.0626}]}\BibitemShut {NoStop}%
\bibitem [{\citenamefont {Petcov}\ and\ \citenamefont
  {Piai}(2002)}]{Petcov:2001sy}%
  \BibitemOpen
  \bibfield  {author} {\bibinfo {author} {\bibfnamefont {S.~T.}\ \bibnamefont
  {Petcov}}\ and\ \bibinfo {author} {\bibfnamefont {M.}~\bibnamefont {Piai}},\
  }\bibfield  {booktitle} {\href {\doibase 10.1016/S0370-2693(02)01591-5}
  {}\bibfield  {journal} {\bibinfo  {journal} {Phys.Lett.}\ }\textbf {\bibinfo
  {volume} {B533}\ }(\bibinfo {year} {2002})\ \bibinfo {pages} {94} [\Eprint
  {http://arxiv.org/abs/hep-ph/0112074} {arXiv:hep-ph/0112074}]}\BibitemShut
  {NoStop}%
\bibitem [{\citenamefont {Choubey}\ \emph {et~al.}(2003)\citenamefont
  {Choubey}, \citenamefont {Petcov},\ and\ \citenamefont
  {Piai}}]{Choubey:2003qx}%
  \BibitemOpen
  \bibfield  {author} {\bibinfo {author} {\bibfnamefont {S.}~\bibnamefont
  {Choubey}}, \bibinfo {author} {\bibfnamefont {S.~T.}\ \bibnamefont {Petcov}}\
  and\ \bibinfo {author} {\bibfnamefont {M.}~\bibnamefont {Piai}},\ }\bibfield
  {booktitle} {\href {\doibase 10.1103/PhysRevD.68.113006} {}\bibfield
  {journal} {\bibinfo  {journal} {Phys.Rev.}\ }\textbf {\bibinfo {volume}
  {D68}\ }(\bibinfo {year} {2003})\ \bibinfo {pages} {113006} [\Eprint
  {http://arxiv.org/abs/hep-ph/0306017} {arXiv:hep-ph/0306017}]}\BibitemShut
  {NoStop}%
\bibitem [{\citenamefont {Learned}\ \emph {et~al.}(2008)\citenamefont
  {Learned}, \citenamefont {Dye}, \citenamefont {Pakvasa},\ and\ \citenamefont
  {Svoboda}}]{Learned:2006wy}%
  \BibitemOpen
  \bibfield  {author} {\bibinfo {author} {\bibfnamefont {J.}~\bibnamefont
  {Learned}}, \bibinfo {author} {\bibfnamefont {S.~T.}\ \bibnamefont {Dye}},
  \bibinfo {author} {\bibfnamefont {S.}~\bibnamefont {Pakvasa}}\ and\ \bibinfo
  {author} {\bibfnamefont {R.~C.}\ \bibnamefont {Svoboda}},\ }\bibfield
  {booktitle} {\href {\doibase 10.1103/PhysRevD.78.071302} {}\bibfield
  {journal} {\bibinfo  {journal} {Phys.Rev.}\ }\textbf {\bibinfo {volume}
  {D78}\ }(\bibinfo {year} {2008})\ \bibinfo {pages} {071302} [\Eprint
  {http://arxiv.org/abs/hep-ex/0612022} {arXiv:hep-ex/0612022}]}\BibitemShut
  {NoStop}%
\bibitem [{\citenamefont {Zhan}\ \emph {et~al.}(2008)\citenamefont {Zhan},
  \citenamefont {Wang}, \citenamefont {Cao},\ and\ \citenamefont
  {Wen}}]{Zhan:2008id}%
  \BibitemOpen
  \bibfield  {author} {\bibinfo {author} {\bibfnamefont {L.}~\bibnamefont
  {Zhan}}, \bibinfo {author} {\bibfnamefont {Y.}~\bibnamefont {Wang}}, \bibinfo
  {author} {\bibfnamefont {J.}~\bibnamefont {Cao}}\ and\ \bibinfo {author}
  {\bibfnamefont {L.}~\bibnamefont {Wen}},\ }\bibfield  {booktitle} {\href
  {\doibase 10.1103/PhysRevD.78.111103} {}\bibfield  {journal} {\bibinfo
  {journal} {Phys.Rev.}\ }\textbf {\bibinfo {volume} {D78}\ }(\bibinfo {year}
  {2008})\ \bibinfo {pages} {111103} [\Eprint {http://arxiv.org/abs/0807.3203}
  {arXiv:0807.3203}]}\BibitemShut {NoStop}%
\bibitem [{\citenamefont {Batygov}\ \emph {et~al.}(2008)\citenamefont
  {Batygov}, \citenamefont {Dye}, \citenamefont {Learned}, \citenamefont
  {Matsuno}, \citenamefont {Pakvasa},\ and\ \citenamefont
  {Varner}}]{Batygov:2008ku}%
  \BibitemOpen
  \bibfield  {author} {\bibinfo {author} {\bibfnamefont {M.}~\bibnamefont
  {Batygov}}, \bibinfo {author} {\bibfnamefont {S.}~\bibnamefont {Dye}},
  \bibinfo {author} {\bibfnamefont {J.}~\bibnamefont {Learned}}, \bibinfo
  {author} {\bibfnamefont {S.}~\bibnamefont {Matsuno}}, \bibinfo {author}
  {\bibfnamefont {S.}~\bibnamefont {Pakvasa}}\ and\ \bibinfo {author}
  {\bibfnamefont {G.}~\bibnamefont {Varner}},\ }\bibfield  {booktitle}
  {\href@noop {} {}\bibfield  {journal} {\bibinfo  {journal} {\Eprint
  {http://arxiv.org/abs/0810.2580} {arXiv:0810.2580}}\ }(\bibinfo {year}
  {2008})}\BibitemShut {NoStop}%
\bibitem [{\citenamefont {Zhan}\ \emph {et~al.}(2009)\citenamefont {Zhan},
  \citenamefont {Wang}, \citenamefont {Cao},\ and\ \citenamefont
  {Wen}}]{Zhan:2009rs}%
  \BibitemOpen
  \bibfield  {author} {\bibinfo {author} {\bibfnamefont {L.}~\bibnamefont
  {Zhan}}, \bibinfo {author} {\bibfnamefont {Y.}~\bibnamefont {Wang}}, \bibinfo
  {author} {\bibfnamefont {J.}~\bibnamefont {Cao}}\ and\ \bibinfo {author}
  {\bibfnamefont {L.}~\bibnamefont {Wen}},\ }\bibfield  {booktitle} {\href
  {\doibase 10.1103/PhysRevD.79.073007} {}\bibfield  {journal} {\bibinfo
  {journal} {Phys.Rev.}\ }\textbf {\bibinfo {volume} {D79}\ }(\bibinfo {year}
  {2009})\ \bibinfo {pages} {073007} [\Eprint {http://arxiv.org/abs/0901.2976}
  {arXiv:0901.2976}]}\BibitemShut {NoStop}%
\bibitem [{\citenamefont {Ghoshal}\ and\ \citenamefont
  {Petcov}(2011)}]{Ghoshal:2010wt}%
  \BibitemOpen
  \bibfield  {author} {\bibinfo {author} {\bibfnamefont {P.}~\bibnamefont
  {Ghoshal}}\ and\ \bibinfo {author} {\bibfnamefont {S.~T.}\ \bibnamefont
  {Petcov}},\ }\bibfield  {booktitle} {\href {\doibase 10.1007/JHEP03(2011)058}
  {}\bibfield  {journal} {\bibinfo  {journal} {JHEP}\ }\textbf {\bibinfo
  {volume} {1103}\ }(\bibinfo {year} {2011})\ \bibinfo {pages} {058} [\Eprint
  {http://arxiv.org/abs/1011.1646} {arXiv:1011.1646}]}\BibitemShut {NoStop}%
\bibitem [{\citenamefont {Ciuffoli}\ \emph
  {et~al.}(2012{\natexlab{a}})\citenamefont {Ciuffoli}, \citenamefont
  {Evslin},\ and\ \citenamefont {Zhang}}]{Ciuffoli:2012iz}%
  \BibitemOpen
  \bibfield  {author} {\bibinfo {author} {\bibfnamefont {E.}~\bibnamefont
  {Ciuffoli}}, \bibinfo {author} {\bibfnamefont {J.}~\bibnamefont {Evslin}}\
  and\ \bibinfo {author} {\bibfnamefont {X.}~\bibnamefont {Zhang}},\ }\bibfield
   {booktitle} {\href@noop {} {}\bibfield  {journal} {\bibinfo  {journal}
  {\Eprint {http://arxiv.org/abs/1208.1991} {arXiv:1208.1991}}\ }(\bibinfo
  {year} {2012}{\natexlab{a}})}\BibitemShut {NoStop}%
\bibitem [{\citenamefont {Ciuffoli}\ \emph
  {et~al.}(2012{\natexlab{b}})\citenamefont {Ciuffoli}, \citenamefont
  {Evslin},\ and\ \citenamefont {Zhang}}]{Ciuffoli:2012bs}%
  \BibitemOpen
  \bibfield  {author} {\bibinfo {author} {\bibfnamefont {E.}~\bibnamefont
  {Ciuffoli}}, \bibinfo {author} {\bibfnamefont {J.}~\bibnamefont {Evslin}}\
  and\ \bibinfo {author} {\bibfnamefont {X.}~\bibnamefont {Zhang}},\ }\bibfield
   {booktitle} {\href@noop {} {}\bibfield  {journal} {\bibinfo  {journal}
  {\Eprint {http://arxiv.org/abs/1209.2227} {arXiv:1209.2227}}\ }(\bibinfo
  {year} {2012}{\natexlab{b}})}\BibitemShut {NoStop}%
\bibitem [{\citenamefont {Qian}\ \emph
  {et~al.}(2012{\natexlab{a}})\citenamefont {Qian}, \citenamefont {Dwyer},
  \citenamefont {McKeown}, \citenamefont {Vogel}, \citenamefont {Wang},\ and\
  \citenamefont {Zhang}}]{Qian:2012xh}%
  \BibitemOpen
  \bibfield  {author} {\bibinfo {author} {\bibfnamefont {X.}~\bibnamefont
  {Qian}}, \bibinfo {author} {\bibfnamefont {D.~A.}\ \bibnamefont {Dwyer}},
  \bibinfo {author} {\bibfnamefont {R.~D.}\ \bibnamefont {McKeown}}, \bibinfo
  {author} {\bibfnamefont {P.}~\bibnamefont {Vogel}}, \bibinfo {author}
  {\bibfnamefont {W.}~\bibnamefont {Wang}}\ and\ \bibinfo {author}
  {\bibfnamefont {C.}~\bibnamefont {Zhang}},\ }\bibfield  {booktitle}
  {\href@noop {} {}\bibfield  {journal} {\bibinfo  {journal} {\Eprint
  {http://arxiv.org/abs/1208.1551} {arXiv:1208.1551}}\ }(\bibinfo {year}
  {2012}{\natexlab{a}})}\BibitemShut {NoStop}%
\bibitem [{\citenamefont {Ghoshal}\ and\ \citenamefont
  {Petcov}(2012)}]{Ghoshal:2012ju}%
  \BibitemOpen
  \bibfield  {author} {\bibinfo {author} {\bibfnamefont {P.}~\bibnamefont
  {Ghoshal}}\ and\ \bibinfo {author} {\bibfnamefont {S.~T.}\ \bibnamefont
  {Petcov}},\ }\bibfield  {booktitle} {\href@noop {} {}\bibfield  {journal}
  {\bibinfo  {journal} {\Eprint {http://arxiv.org/abs/1208.6473}
  {arXiv:1208.6473}}\ }(\bibinfo {year} {2012})}\BibitemShut {NoStop}%
\bibitem [{\citenamefont {Li}\ \emph {et~al.}(2013)\citenamefont {Li},
  \citenamefont {Cao}, \citenamefont {Wang},\ and\ \citenamefont
  {Zhan}}]{Li:2013zyd}%
  \BibitemOpen
  \bibfield  {author} {\bibinfo {author} {\bibfnamefont {Y.-F.}\ \bibnamefont
  {Li}}, \bibinfo {author} {\bibfnamefont {J.}~\bibnamefont {Cao}}, \bibinfo
  {author} {\bibfnamefont {Y.}~\bibnamefont {Wang}}\ and\ \bibinfo {author}
  {\bibfnamefont {L.}~\bibnamefont {Zhan}},\ }\bibfield  {booktitle}
  {\href@noop {} {}\bibfield  {journal} {\bibinfo  {journal} {\Eprint
  {http://arxiv.org/abs/1303.6733} {arXiv:1303.6733}}\ }(\bibinfo {year}
  {2013})}\BibitemShut {NoStop}%
\bibitem [{\citenamefont {Qian}\ \emph
  {et~al.}(2012{\natexlab{b}})\citenamefont {Qian}, \citenamefont {Tan},
  \citenamefont {Wang}, \citenamefont {Ling}, \citenamefont {McKeown},\ and\
  \citenamefont {Zhang}}]{Qian:2012zn}%
  \BibitemOpen
  \bibfield  {author} {\bibinfo {author} {\bibfnamefont {X.}~\bibnamefont
  {Qian}}, \bibinfo {author} {\bibfnamefont {A.}~\bibnamefont {Tan}}, \bibinfo
  {author} {\bibfnamefont {W.}~\bibnamefont {Wang}}, \bibinfo {author}
  {\bibfnamefont {J.~J.}\ \bibnamefont {Ling}}, \bibinfo {author}
  {\bibfnamefont {R.~D.}\ \bibnamefont {McKeown}}\ and\ \bibinfo {author}
  {\bibfnamefont {C.}~\bibnamefont {Zhang}},\ }\bibfield  {booktitle}
  {\href@noop {} {}\bibfield  {journal} {\bibinfo  {journal} {\Eprint
  {http://arxiv.org/abs/1210.3651} {arXiv:1210.3651}}\ }(\bibinfo {year}
  {2012}{\natexlab{b}})}\BibitemShut {NoStop}%
\bibitem [{\citenamefont {Huber}\ and\ \citenamefont
  {Schwetz}(2004)}]{Huber:2004xh}%
  \BibitemOpen
  \bibfield  {author} {\bibinfo {author} {\bibfnamefont {P.}~\bibnamefont
  {Huber}}\ and\ \bibinfo {author} {\bibfnamefont {T.}~\bibnamefont
  {Schwetz}},\ }\bibfield  {booktitle} {\href {\doibase
  10.1103/PhysRevD.70.053011} {}\bibfield  {journal} {\bibinfo  {journal}
  {Phys.Rev.}\ }\textbf {\bibinfo {volume} {D70}\ }(\bibinfo {year} {2004})\
  \bibinfo {pages} {053011} [\Eprint {http://arxiv.org/abs/hep-ph/0407026}
  {arXiv:hep-ph/0407026}]}\BibitemShut {NoStop}%
\bibitem [{\citenamefont {Vogel}\ and\ \citenamefont
  {Engel}(1989)}]{Vogel:1989iv}%
  \BibitemOpen
  \bibfield  {author} {\bibinfo {author} {\bibfnamefont {P.}~\bibnamefont
  {Vogel}}\ and\ \bibinfo {author} {\bibfnamefont {J.}~\bibnamefont {Engel}},\
  }\bibfield  {booktitle} {\href {\doibase 10.1103/PhysRevD.39.3378}
  {}\bibfield  {journal} {\bibinfo  {journal} {Phys.Rev.}\ }\textbf {\bibinfo
  {volume} {D39}\ }(\bibinfo {year} {1989})\ \bibinfo {pages}
  {3378}}\BibitemShut {NoStop}%
\bibitem [{\citenamefont {Hagiwara}\ \emph {et~al.}(2011)\citenamefont
  {Hagiwara}, \citenamefont {Okamura},\ and\ \citenamefont
  {Senda}}]{Hagiwara:2011kw}%
  \BibitemOpen
  \bibfield  {author} {\bibinfo {author} {\bibfnamefont {K.}~\bibnamefont
  {Hagiwara}}, \bibinfo {author} {\bibfnamefont {N.}~\bibnamefont {Okamura}}\
  and\ \bibinfo {author} {\bibfnamefont {K.}~\bibnamefont {Senda}},\ }\bibfield
   {booktitle} {\href {\doibase 10.1007/JHEP09(2011)082} {}\bibfield  {journal}
  {\bibinfo  {journal} {JHEP}\ }\textbf {\bibinfo {volume} {1109}\ }(\bibinfo
  {year} {2011})\ \bibinfo {pages} {082} [\Eprint
  {http://arxiv.org/abs/1107.5857} {arXiv:1107.5857}]}\BibitemShut {NoStop}%
\bibitem [{\citenamefont {Abe}\ \emph {et~al.}(2012)\citenamefont {Abe} \emph
  {et~al.}}]{Abe:2011fz}%
  \BibitemOpen
  \bibfield  {author} {\bibinfo {author} {\bibfnamefont {Y.}~\bibnamefont
  {Abe}} \emph {et~al.} (\bibinfo {collaboration} {DOUBLE-CHOOZ
  Collaboration}),\ }\bibfield  {booktitle} {\href {\doibase
  10.1103/PhysRevLett.108.131801} {}\bibfield  {journal} {\bibinfo  {journal}
  {Phys.Rev.Lett.}\ }\textbf {\bibinfo {volume} {108}\ }(\bibinfo {year}
  {2012})\ \bibinfo {pages} {131801} [\Eprint {http://arxiv.org/abs/1112.6353}
  {arXiv:1112.6353}]}\BibitemShut {NoStop}%
\bibitem [{\citenamefont {Vogel}\ and\ \citenamefont
  {Beacom}(1999)}]{Vogel:1999zy}%
  \BibitemOpen
  \bibfield  {author} {\bibinfo {author} {\bibfnamefont {P.}~\bibnamefont
  {Vogel}}\ and\ \bibinfo {author} {\bibfnamefont {J.~F.}\ \bibnamefont
  {Beacom}},\ }\bibfield  {booktitle} {\href {\doibase
  10.1103/PhysRevD.60.053003} {}\bibfield  {journal} {\bibinfo  {journal}
  {Phys.Rev.}\ }\textbf {\bibinfo {volume} {D60}\ }(\bibinfo {year} {1999})\
  \bibinfo {pages} {053003} [\Eprint {http://arxiv.org/abs/hep-ph/9903554}
  {arXiv:hep-ph/9903554}]}\BibitemShut {NoStop}%
\bibitem [{Ere()}]{Eres}%
  \BibitemOpen
  \bibfield  {author} {\href@noop {} {}}\bibinfo {note} {X. Qian and Y. Wang,
  private communication.}\BibitemShut {Stop}%
\bibitem [{\citenamefont {Beringer}\ \emph {et~al.}(2012)\citenamefont
  {Beringer} \emph {et~al.}}]{Beringer:1900zz}%
  \BibitemOpen
  \bibfield  {author} {\bibinfo {author} {\bibfnamefont {J.}~\bibnamefont
  {Beringer}} \emph {et~al.} (\bibinfo {collaboration} {Particle Data Group}),\
  }\bibfield  {booktitle} {\href {\doibase 10.1103/PhysRevD.86.010001}
  {}\bibfield  {journal} {\bibinfo  {journal} {Phys.Rev.}\ }\textbf {\bibinfo
  {volume} {D86}\ }(\bibinfo {year} {2012})\ \bibinfo {pages}
  {010001}}\BibitemShut {NoStop}%
\bibitem [{Err()}]{Error_dmm31}%
  \BibitemOpen
  \bibfield  {author} {\href@noop {} {}}\bibinfo {note} {J. Cao, talk at
  ICHEP2012 in Melbourne.}\BibitemShut {Stop}%
\end{thebibliography}%

\end{document}